# Enhancing Grid Resilience for Giga-Watt Scale Data Centers Using High Voltage Circuit Breaker Operated Braking Resistors

Soham Ghosh[1] and Mohammad Ashraf Hossain Sadi[2]
[1]Black & Veatch, Overland Park, Kansas 66211, USA, Email: sghosh27@ieee.org
[2]University of Central Missouri, Warrensburg, Missouri 64093, USA, Email: sadi@ucmo.edu

*Abstract*—As hyperscale and co-located data centers scale, the electric grid sees an increase in large, voltage-sensitive IT loads with these data center plant size ranging between 500 MW to 2 GW. A sudden loss of these loads as they switch to onsite UPS during grid voltage excursion events causes a grid frequency rise from generation and load imbalance, and a voltage rise because less power is flowing through the network. This paper proposes and theoretically demonstrates the use of high voltage circuit breaker operated braking resistors at data center transmission substations as an effective strategy in enhancing grid resilience under such large load loss scenarios. We developed a test bed to illustrate the dynamic behavior of the system with resistive braking on a gigawatt scale data center load cluster connected to a 345 kV network. The braking resistor(s), which in the case of inverter rich system comes in a multi-stage configuration, are connected/disconnected via high-speed circuit breaker(s). Results show that insertion for 0.25 – 0.85 seconds sufficiently reduce rate of change of frequency and provides time for primary governor response and capacitor switching to restore steady state. Sensitivity across different synchronous machines/ inverter-based resource mix are tested and confirms robustness. We conclude circuit breaker controlled resistive braking is a practical means to enhance Bulk Electric System (BES) resilience for gigawatt scale data centers. The approach integrates with protection, needs no generator changes, and can be scaled with cluster size/growth of the data center facility load.

## I. Introduction

Since 2022, with the world entering the post-covid era, the data center industry has experienced a sharp growth, driven by the confluence of technological advancements like AI and 5G, the expansion of cloud computing and edge applications, and the increasing demand for data storage and processing. This boom is evident in the rapid growth of hyperscale, colocation, and edge data centers. AI data centers can be broadly categorized into hyperscale, colocation, edge, and cloud data centers, each offering unique characteristics and catering to different needs in the AI landscape. Hyperscale centers focus on massive scale and efficiency for large-scale AI training, while colocation centers provide shared infrastructure for various organizations. Edge data centers, on the other hand, are smaller and closer to end-users, optimized for low-latency applications like real-time AI inference. Cloud data centers, like those offered by AWS, Azure, or Google Cloud, provide on-demand AI services and infrastructure through a subscription model. Compounded with this rapid growth, these data center facilities are now projected to consume energy in the giga-watt scale. Estimating the typical trending wattage usage for advanced data centers per query is complex, as it varies significantly based on numerous factors. However, some estimates indicate that a single query to a large language model (LLM) like ChatGPT can consume around 2.9 watt-hours of electricity, compared to roughly 0.3 watt-hours for a traditional Google search query. This signifies a substantially higher energy demand for AI-driven queries, often estimated to be ten times or more compared to standard web searches. From a data center power delivery standpoint pioneering work being conducted can be categorized into three broad fronts:

1. Research being conducted to ensure more homogenous task scheduling [1] and allowing a higher degree of proactive work load shift [2],

2. Work being done to find ways that would allow commissioning and powering up these new data centers, while navigating the interconnection obligation process with transmission system operator and distribution system operator (TSO/DSO) or using islanded/behind the meter generation [3, 4],

3. Industry task force [5, 6] working groups are conducting system-level studies to estimate the potential scale of load loss during system disturbances, while also evaluating strategies to mitigate and reduce the impact of such large-scale load interruptions on overall system reliability.

The third category is particularly interesting from a long term impact standpoint, given actions taken by transmission planners and operators based on the recommendation of the industry task force and working groups will pave the utility policies of tomorrow. Our work aims to contribute towards this niche third category of collective industry effort to minimize system impact with the loss of large loads in the gigawatt scale. It is important to note here that the definition of large load can vary between TSOs; for instance within the Electric Reliability Council of Texas (ERCOT) footprint large load is anything above 75 MW, while in the New York Independent System Operator (NYISO) footprint the definition is indirect; a load of 10 MW or larger at 115 kV and higher or a load 80 MW or greater connected below 115 kV is be subjected to NYISO's interconnection study requirements [5]. The motivation for our work and the contributions of this manuscript are outlined in the next two sections.

*A. Research motivation*

In January of 2025, the North American Electric Reliability Corporation (NERC) published an incident review report [7] making bulk electric system operators cognizant of the risks and operational challenges posed by the rapid integration of large, voltage-sensitive loads into the grid. In one particular incident from July 10, 2024, that the report highlighted, a shunted surge arrestor on a 230 kV transmission line triggered multi-shot automatic recloser operation, causing six recordable voltage depression, from back-to-back reclose operation at the two ends of the transmission line. The voltage dips triggered customer-driven disconnection of roughly 1,500 MW of voltage-sensitive demand, an event not foreseen by bulk system operators. The resulting load loss caused both frequency and voltage to increase, with frequency peaking at 60.047 Hz before stabilizing near 60.0 Hz within about four minutes. Voltage reached a maximum of 1.07 per unit, after which operators withdrew shunt capacitor banks in the affected region to restore levels to their normal operating range. The scale of this load loss motivated us to study how a test bed system would react if the load loss can be staggered in time by using one or multiple load-side braking resistor, connected to a major transmission level substation serving these data center loads.

*B. Manuscript contribution*

This study advances the application of braking resistors which has traditionally been used on the generation side [8] to increase inter-area maximum power transfer level which is otherwise constrained by transient stability limit. These braking resistors had allowed the transfer of additional power while ensuring the generator(s) doesn't lose synchronism in the event of severe bolted three phase faults. With data center loads growing in size and being spatio-concentrated, one can take the advantage of placing braking resistor in major transmission substation that serves these data center loads. To this effect the key contribution of this manuscript are as follows:

*1)* We categorize the different load types that constitutes a data center load and provide recommendation on protective relaying configuration such that non-voltage sensistive loads remain connected to the grid during multi-shot reclosing induced voltage excursion events.

*2)* We develop a test bed system to observe the dynamic performance of a synchronous generator's power output in response to a loss in data center IT load with varying single-stage resistive break sizes.

*3)* The test bed is extended to represent different synchronous machine and inverter based generation compositions. Systems with high proportions of inverter based resources are tested with multi-stage resistive braking.

*4)* An alternate approach to time domain simulation is proposed to evaluate system stability by tracing the Eigen value trajectory of the system under two different short circuit ratios (SCR) and varying resistive braking loads.

The use of traditional thyristor controlled resistive breaks were originally intended to improve the maximum power transfer level in systems constrained by the transient stability limit. In most braking resistor applications, the 'ON–OFF' thyristor control approach was typically governed by local monitoring of the generator's real power output and the rate of change of that output [9]. The control law for these applications had to be precise as excessive retardation of the machine detoriated the system performance, and hence thyristor controlled switching had an advantage over mechanical switching for these use cases.

However, for frequency stabilization for data center IT load load, as we shall see, the size of the braking resistors are set smaller than the IT load lost, and as such there is no risk for excessive retardation of the machine speed. Hence the timing for which the resistors may remain online need not be precisely controlled and are generally limited by the thermal and mechanical strength of the composite stainless steel stranded resistor conductors. Instead of using traditional thyristor controlled resistive breaks which tends to be significantly more expensive, we demonstrate that single stage or multi-stage high voltage circuit breaker operated resistive break units can achieve the desired result of frequency stabilization due to data center load dropping from the grid at a lower cost.

The remainder of the manuscript is structured as follows: section II provides a theoretical background on the application of braking resistor for data centers. This section also provides guidelines on how protective relaying schemes can be designed to ensure only voltage-sensitive IT loads are switched to UPS during voltage disturbances. Section III talks about the modelling aspects of the test bed system. Section IV provides detailed discussion on the behavior of the system from a stability standpoint under various test scenarios. Finally, Section V summarizes the key findings, outlines current limitations of the work, and identifies directions for future research.

## II. NON-VOLTAGE SENSISTIVE LOAD RIDE THROUGH AND THOUGHTS ON USING BRAKING RESISTOR

Data center electrical demand is dominated by three hardware classes [10, 11]:

1. Information technology (IT) equipment – These are voltage and frequency sensitive and must operate within the ITIC/CBEMA power-quality envelope. These loads typically make up 40 to 50 percent of total load, and includes servers, storage arrays, and network devices such as switches, routers, and firewalls.

2. Thermal management systems – These loads are relatively tolerant to millisecond level voltage and frequency excursions. They make up approximately 30 to 40 percent of the total load, spanning conventional HVAC and advanced liquid-cooling systems.

3. Auxiliary and balance-of-plant loads - These are generally non-sensitive loads that make up about 10 to 15 percent of the total load composition, and includes building lighting, auxiliary pumps, and security systems.

To ensure that only voltage sensitive IT loads are switched to on-site UPS and non-voltage sensitive HVAC and auxiliary loads remain connected to the electric grid during voltage excursion event, it is important to understand the power delivery and protection architecture of the data center. A representative 345 kV/ 34.5 kV/480 V data center power delivery architecture is shown in Figure 1. The accompanying 345/34.5 kV transmission substation feeds five data center buildings, each with a 200 MW rating (hence a total load of 1 GW). The IT feeders downstream are designed to have its dedicated voltage relays performing ANSI 27/59 functions in concurrence to the ITIC/CBEMA power-quality envelope. For grid induced voltage excursions, the voltage sensitive IT loads can be switched to UPS, while the HVAC and other auxiliary loads remains connected to the grid. Other protection functions such as aggregated overcurrent protection (ANSI 50/51) of IT and non-IT load can be provided at the plant level, which would allow downstream overcurrent protection of the plant in the event of plat level fault. The benefit of this hierarchical level of protection can be illustrated through two scenarios.

*Scenario 1* – If a fault occurs on the 480 V bus at data center 1 (Figure 1, fault location 1), the bus protection scheme isolates the faulted section and transfers the entire downstream site load, including both IT and non-IT, to the UPS. With a total data center 1 load of 200 MW, the grid sees a 200 MW reduction in demand as a result of the bus overcurrent protection operation. With such a plant level fault at data center 1, data center 2 – 5 remains unaffected.

*Scenario 2* – Assuming a transmission line surge arrester malfunctions (Figure 1, fault location 2), a multi-shot reclosing sequence is typically initiated. The resulting momentary voltage sags is picked up by ANSI 27/59 relays at data centers 1 through 5, causing sensitive IT loads to transfer to on-site UPS systems, while non-IT loads remain connected to the grid. If each data center is 200 MW with a 1:1 IT to non-IT load split, the maximum load drop at the interconnected transmission substation is going to be 5 × 100 MW = 500 MW.

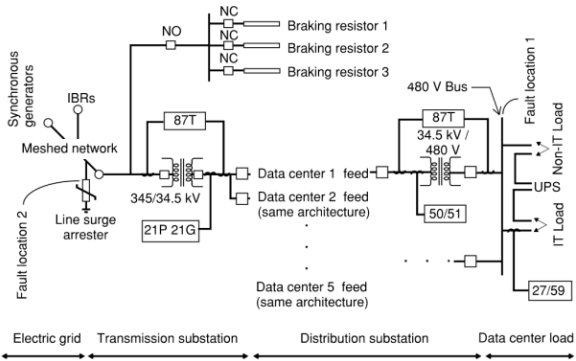

Fig. 1. A conceptual single line power delivery architecture for a data center with hierarchical level of protection.

If the power delivery architecture for the gigawatt scale data center load cluster is set similar to Figure 1, between scenarios 1 and 2, one may conclude that the worst case load loss that the interconnected transmission substation is likely to witness is 500 MW, which provides an initial ballpark for sizing the braking resistor. The mathematical theory on the desired sizing of the braking resistor(s) need to be further developed. For such an illustration, let the per unit speed deviation be $\omega'(t)$. For a step loss of load the one-machine swing equation with a shunt braking resistor can be written as shown in (1), with $P_{br}$ being the real power absorbed by the braking resistor while it is in service, and $\Delta P$ denoting the tripped voltage sensitive IT load (recall this is 500 MW from illustration of scenario 2).

$$2H \frac{d\omega'}{dt} = \Delta P - P_{br} - D\omega' \qquad (1)$$

If the braking resistor is connected to a bus whose voltage stays near $V_{pu}$, then the real power absorbed by the braking resistor, $P_{br}$, can be computed using (2).

$$P_{br} \approx \frac{V_{pu}^2}{R_{br,pu}} \qquad (2)$$

Suppose the resistor is switched in at $t = 0$, with an initial speed deviation $\omega'(0) = \omega'_0$, the closed form solution while the resistor is kept online can be written via equation (3). At this point, one may choose a removal criterion, for example $\omega'(T) = \omega'_{target}$, and solve for $T$, which should yield (4).

$$\omega'(t) = \left(\omega'_0 - \frac{\Delta P - P_{br}}{D}\right)e^{-\frac{D}{2H}t} + \frac{\Delta P - P_{br}}{D} \qquad (3)$$

$$T = \frac{2H}{D} \ln\left(\frac{\omega'_0 - \frac{\Delta P - P_{br}}{D}}{\omega'_{target} - \frac{\Delta P - P_{br}}{D}}\right) \qquad (4)$$

With a common first swing approximation of $D \approx 0$, the time ($T$) required to bring $\omega'$ back to $\omega'_{target}$ becomes as shown in (5).

$$T = \frac{2H}{P_{br} - \Delta P}(\omega'_0 - \omega'_{target}) \qquad (5)$$

It is interesting to note that without intervention of primary governor response, $P_{br} > \Delta P$ yields reverse acceleration while $P_{br} \leq \Delta P$ only reduces the acceleration. The case of $P_{br} \leq \Delta P$ will be developed in the subsequent section as this leverages the governor response of the generators and results in a braking resistor of size smaller or equal to the maximum voltage sensitive IT load of the aggregated data center cluster. Performance evaluations shall be made for single and multiple stage resistive braking loads of size smaller than the maximum voltage sensitive IT load of the plant.

### III. TEST BED SYSTEM DESCRIPTION

The effectiveness of the proposed controlled braking resistors at data center transmission substation in enhancing grid resilience is examined and verified in a MATLAB/SIMULINK model as presented in Figure 1. The grid side synchronous generator (SG) and the inverter based resources (IBRs) are designed to connect to a meshed network. For the SG, we have used an IEEE type 1 excitation and governor system [12]. The braking resistor(s) are placed in the point of common coupling (PCC) position of the transmission substation, and they are controlled using high voltage circuit breakers as previously proposed in the manuscript. The IBR is designed as a three-phase grid-connected grid forming (GFM) inverter with properly designed inverter side filter inductance and capacitance. In this work we choose grid forming inverter (GFM) over the grid following inverter (GFL) as the IBR because the GFL inverters often face stability issues in weak grids [13]. In contrast, grid-forming (GFM) inverters can be controlled and operate stably against weak grids since they are controlled as voltage sources and independently regulate their output voltages and frequencies [14]. Main parameters of the SG, IBR, and the meshed network are provided in Table I for purpose of reproducibility of the results.

TABLE I. MATLAB SIMULATION PARAMETERS

| System | Description | Value |
|---|---|---|
| Generator | Rated Power (MW) | 500 |
| | Armature resistance $r_a$ (p.u.), reactance $x_a$ (p.u.) | 0.003, 0.102 |
| | Inertia, H (sec) | 11 |
| GFM | Rated Power (MW) | 500 |
| | Filter inductance/capacitance | 3 mH/30 μF |
| | Short circuit capability | 1.3 p.u. |
| Line Parameters | Resistance per km | $1 \times 10^{-4}$ |
| | Inductance per km | $1 \times 10^{-3}$ |

## IV. SIMULATION RESULTS AND DISCUSSION

In this section, the system is incrementally developed from a single machine system with single-stage resistive braking setup to more involved multi-machine systems and multi-stage resistive braking scenarios.

### A. Single machine system with single-stage resistive braking

With a test bed system developed as shown in Figure 1, and system parameters as described in section III, the first set of simulations were conducted on a setup consisting of a single synchronous machine as a generation source. The data center load consists of a cumulative 1 GW of load spread across five data center buildings. With a 1:1 IT to non-IT load split, the total IT load based on the aggregation of the five data center buildings is 500 MW.

Simulation was initiated at $t = 0$ s and reaches steady state within a few cycles. At $t = 0.1$ s, 500 MW of IT loads was dropped and switched to on-site UPS thereby mimicking a grid-induced voltage excursion event. With a time delay of 3 cycles

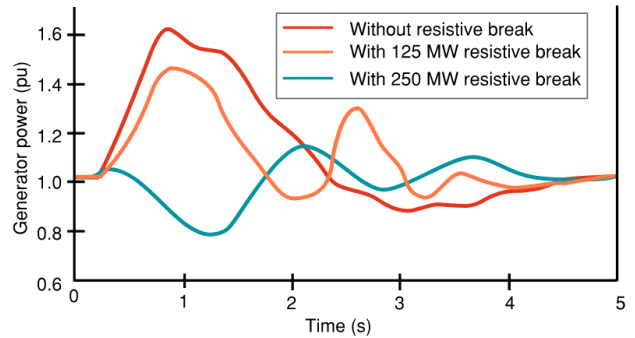

Fig. 2. Synchronous generator's output power dynamics with a single stage resistive braking load.

(1 for relay pickup and 2 for breaker operation) a resistive braking load was introduced to arrest the power of the synchronous generator. The dynamical of the generator's power with varying sizes of the resistive braking load were plotted in Figure 2; with the resistive braking load kept online for the remaining duration of the simulation. As one may observe, without any resistive braking load, there is a sharp rise in the generator's power reaching 1.6 p.u., which can cause the generator to be tripped offline. With incrementally larger braking resistors (125 MW and 250 MW), the jump in the power of the generator is arrested. Given the cumulative IT load of the plant is 500 MW, the largest single stage resistive braking load tested on the system is 250 MW, which is half the size of the plant's IT load. Higher single stage resistive braking load may better arrest the initial power rise when the brake is brought online, but will cause a larger generator power surge upon disconnection, and hence not recommended.

### B. Multi-machine system with single-stage resistive braking

Given modern interconnected power system are witnessing higher penetration of renewable resources, it is imperative to investigate the response of such a system in wake of a large load loss. To accomplish this, multiple scenarios were analyzed with different proportions of traditional synchronous machine based generations and inverter based resources. Similar to the previous simulation setup, at $t = 0.1$ s 500 MW of IT loads were dropped and switched to on-site UPS thereby mimicking a grid-induced voltage excursion event, and a time delay of 3 cycles (1 for relay pickup and 2 for breaker operation) was given prior to bringing a 250 MW resistive braking load online. At 0.25 s the single stage resistive break is switched offline by tripping the connecting circuit breaker. The dynamics of the system in terms of system frequency, for three different generation mixes are captured, with results shown in Figure 3. A higher proportion of IBR in the system leads to more sustained frequency rise and frequency oscillation. The worst case scenario, with 25% synchronous machines and 75% IBRs, will be developed further with multi-stage resistive breaks, to improve system frequency dynamics.

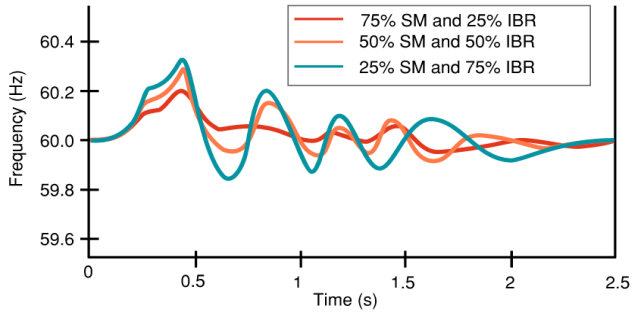

Fig. 3. System frequency with different SM/IBR combinations and a single stage resistive braking load.

## C. IBR rich system with multi-stage resistive braking

For a system with a high penetration of IBRs, the system composition was set at 25% synchronous machine and 75% inverter-based resource. Instead of using a single stage 250 MW braking resistor, the load transition steps were lowered by using a multi-stage resistive braking system. Stages 1, 2, and 3 of this multi-stage resistor bank was composed of individual 130 MW, 130 MW, and 110 MW resistor respectively, thereby decreasing the load step change from 250 MW to a maximum of 130 MW, thereby improving the frequency transients.

For the simulation, multiple resistor-switching timings were evaluated with an objective to limit both the time for which the resistors had to be kept online and minimizing the frequency transients. The initial steps of the simulation were kept the same, with the data center IT load dropping out at $t = 0.1$ s, and the 370 MW multi-stage resistor bank coming online in 3 cycles. Resistor 1 (130 MW) was dropped via a circuit breaker trip command after being online for 0.1 second, resistor 2 (130 MW) was dropped 0.25 second after resistor 1 was dropped, and resistor 3 (110 MW) was dropped 0.5 seconds after resistor 2 was dropped. Figure 4 shows the simulation comparison of a high penetration IBR system, comparing the system's frequency dynamics for single and multi-stage resistive braking. As one may infer, an IBR rich system that is inherently sensitive to load changes shall benefit from a multi-stage braking solution.

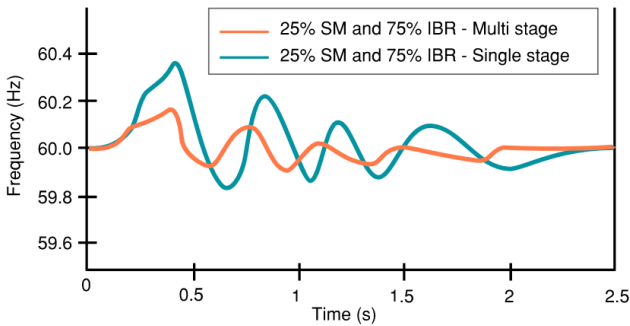

Fig. 4. System frequency comparison for a system with 25%SM/ 75% IBR with single-stage versus multi-stage resistive braking load.

For this last study, the total duration for which any of the resistors are kept online was limited to 1 second or less. Assessment of the upper bound of this time limit for which the resistors can be kept online is important to assess in order to prevent damage of the resistors' multi-strand stainless-steel wires from thermal heat dissipation and electromagnetic tension forces. The thickness of the multi-stranded stainless-steel resistor wires should be selected such that they can withstand the nominal load ampacity induced temperate rise till the upper bound of the time limit with some additional margin to factor in primary breaker failure and tripping through backup breaker operation.

## D. System stability evaluation through eigenvalue trajectories

As an alternate to the time domain simulation assessment of system dynamics for a given system with a certain generation composition and multi-stage resistive braking setup, tracing the eigen value trajectory of the system may be an alternate approach. From classical control theory, eigenvalues reveal the small-signal stability of a linear system by indicating whether its dynamics decay to a stable state or diverge. A system is considered stable when every eigenvalue lies in the left half of the complex plane. The presence of any eigenvalue with a positive real component indicates system instability.

Figure 5 represent the dominant eigenvalue anaysis under two different short circuit ratios (SCR) with varying resistive braking loads. For the weak grid (SCR 2.0), eigenvalues tends to approach closer to the zero on the real axis as resistive braking load size is reduced, indicating lower damping and higher potential for oscillatory instability. For strong grid (SCR 5.0), eigenvalues shift further left representing stronger damping characteristics. For each SCR conditions, the spread out eigenvalue clusters represent how different network impedance characteristics affect the stability margins.

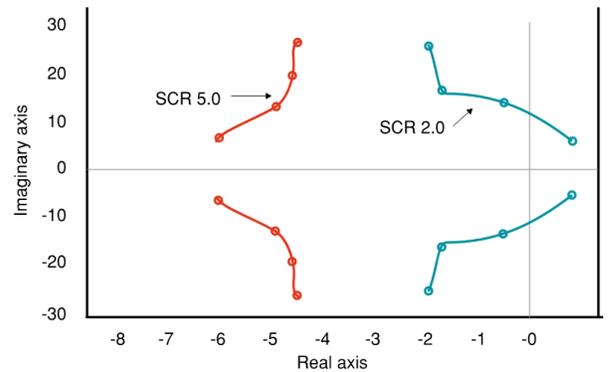

Fig. 5. Eigenvalue trajectories for SCR of 2.0 and 5.0 with varying resistive braking load.

## E. Investigating risk of motor stalling due to grid induced voltage disturbances

NERC in the 'considering simultaneous voltage-sensitive load reductions' incident report [7] stated that the voltage magnitude drop during the July 10, 2024 ranged from 0.25 to 0.40 per unit in the load-loss area, with the voltage depression lasting a longest of 66 milliseconds. We conducted supplementary simulations to assess the risk of wide-scale motor stalling, given induction motor from HVAC and cooling systems makes up a major portion of the non-IT data center load. For voltage depressions down to 0.25 per unit and lasting as long as 100 milliseconds, no motor stalling issues were observed within our test-bed system, as can be seen from in Figure 6, which reinforces the initial assumption that during such grid induced voltage excursion event, the non-IT loads may be kept connected to the grid due to their higher low voltage fault ride through capability.

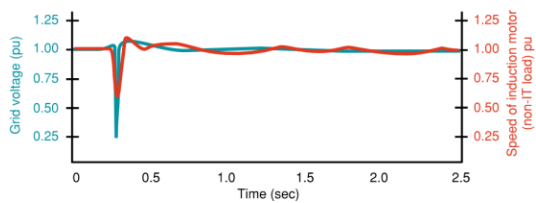

Fig. 6. Investigating risk of largest motor stall - plot of induction motor speed with voltage depressions down to 0.25 per unit and lasting as long as 100 milliseconds.

## V. Conclusion And Future Work

The rapid interconnection of very large data center loads to the bulk power system exposes gaps in current planning and interconnection practices. Many frameworks emphasize static transmission facility ratings and understate system-wide dynamic effects. Because frequency stability depends on interconnection-level inertia and available online generation, and because these characteristics vary widely across the electrical footprint, existing constructs are often insufficient for assessing risk from large load steps.

This paper advances a practical remedy: a first-of-its-kind conceptual application of circuit breaker operated braking resistors to shape net power imbalance and support frequency stability. Test-bed studies show that in IBR-rich systems, multi-stage braking resistors are more effective than a single-stage design. Resistor staging enables tailored energy absorption that moderates rate of change of frequency primary frequency response and other grid controls take effect.

Future work should prioritize thorough evaluations in regions where IBRs operate primarily in grid-following mode, with stress testing across credible contingencies and network conditions. Equally important areas worth exploring are: a) rigorous thermal adequacy and cycling analysis of the resistor hardware to ensure survivability under back-to-back events, and b) real-world pilot validations. Embedding these findings into interconnection studies, specifications, and procurement criteria will allow planners to move beyond static ratings toward frequency-security-aware practice, creating a clear path to safely accommodate emerging large loads without compromising BPS stability.

Acknowledgment and Declaration of Interest

The authors would like to thank M. L Shelton, P. F Winkelman, W. A. Mittalstadt, and W. J. Bellerby for their pioneering 1974 work in designing, constructing, and testing a 1400 MW resistor break at Chief Joseph substation at north central Washington, which formed a motivation for this work.

The views and conclusions expressed are solely those of the authors and do not necessarily reflect the views of their employers or institutions. This work was conducted independently and received no external funding.